\relax
\documentclass[letterpaper]{article} 
\usepackage[utf8]{inputenc}
\usepackage[T1]{fontenc}
\usepackage{aaai18}  
\usepackage{times}  
\usepackage{helvet}  
\usepackage{courier}  
\usepackage{url}  
\usepackage{graphicx}  

\begin{document}
\title{\textit{Mater certa est, pater numquam}:\\
What can Facebook Advertising Data Tell Us about Male Fertility Rates?\thanks{Please cite the version from Proceedings of the Twelfth International Conference on Web and Social Media (ICWSM-2018).}}

\author{
Francesco Rampazzo\textsuperscript{1},\hspace{1.5mm} Emilio Zagheni\textsuperscript{2},\hspace{1.5mm} Ingmar Weber\textsuperscript{3},\hspace{1.5mm}  Maria Rita Testa\textsuperscript{4},\hspace{1.5mm} Francesco Billari\textsuperscript{5}
\\
\textsuperscript{1}University of Southampton \& Max Planck Institute for Demographic Research,\\
\textsuperscript{2}University of Washington \&  Max Planck Institute for Demographic Research,\\
\textsuperscript{3}Qatar Computing Research Institute,
\textsuperscript{4}Wittgenstein Centre,
\textsuperscript{5}Università Bocconi\\
f.rampazzo@soton.ac.uk, emilioz@uw.edu, iweber@hbku.edu.qa, maria.rita.testa@oeaw.ac.at, francesco.billari@unibocconi.it
}


\maketitle

\begin{abstract}

In many developing countries, timely and accurate information about birth rates and other demographic indicators is still lacking, especially for male fertility rates. Using anonymous and aggregate data from Facebook's Advertising Platform, we produce global estimates of the Mean Age at Childbearing (MAC), a key indicator of fertility postponement.  Our analysis indicates that fertility measures based on Facebook data are highly correlated with conventional indicators based on traditional data, for those countries for which we have statistics. For instance, the correlation of the MAC computed using Facebook and United Nations data is 0.47 ($p = 4.02e-08$) and 0.79 ($p = 2.2e-15$) for female and male respectively. Out of sample validation for a simple regression model indicates that the mean absolute percentage error is 2.3\%. We use the linear model and Facebook data to produce estimates of the male MAC for countries for which we do not have data.

\end{abstract}

%
%
\section{Introduction} 
To count and project populations for the world, or a given country, we need timely and accurate information about birth and death rates. This information exists for developed countries, but it is missing for many developing countries that do not have adequate vital registration systems \cite{bongaarts_estimating_2015}. As a result, estimates are often obtained using indirect methods applied to survey data, especially for developing countries and male fertility \cite{un_2015,un_2017}.
The lack of up-to-date information has severe repercussions in the implementation and monitoring of policies. 

This paper aims to understand whether data from Facebook's Advertising platform can be used to obtain estimates of the mean age at childbearing (MAC). The MAC is an important descriptive measure in demographic research, which can help with understanding patterns in fertility behaviour, such as postponement of childbearing. For this, we investigate the extent to which the MAC produced using the Facebook Advertising Platform data are congruent with figures from the United Nations. The significance of the study is to determine whether Facebook is a viable data source for studying two struggling areas of demographic research: male fertility which has been thus far neglected, and the fertility of developing  countries which is hampered by the lack of accurate data \cite{schoumaker_measuring_2017}.

%
%
\section{Fertility Data}

The three main sources of fertility data are (i) vital registration systems, (ii) censuses, and (iii) surveys. These three sources can vary immensely in quality between and within countries and, as Moultrie et al. (\citeyear{moultrie_tools_2013}) suggest, ``\textit{population statistics, like other demographic statistics, whether they are obtained by enumeration, registration, or other means, are subject to error}''.

Vital registration systems have high coverage in developed countries, but not in developing countries. This type of data is generally the key source for estimating births, but in Sub-Saharan Africa they lack representativeness and accuracy \cite{abouzahr_civil_2015}. With improvements in vital registration systems, the percentage of infants whose births had been registered increased from 58\% to 65\% \cite{mikkelsen_global_2015}, but UNICEF (\citeyear{unicef_state_2015}) is still reporting that worldwide one in three infant children (ages 0 to 5) go un-registered.

In countries where vital registrations data is lacking, censuses and surveys are used for estimating births. However, these sources are not timely and accurate: there may be years in between censuses or surveys and lifetime fertility events may be underreported or completely omitted. Indeed, data from these sources in Sub-Saharan Africa are far from accurate \cite{abouzahr_civil_2015}. Outdated surveys and censuses provide fragmented and contradictory information. 
\cite{mikkelsen_global_2015,abouzahr_civil_2015}.

The diffusion of Internet and social media appears to be faster than the improvements made in vital registration.  Although there is need to re-purpose statistical methods for producing statistical inference from Internet data \cite{zagheni_demographic_2015,billari_big_2017}, Facebook can otherwise be thought of as a regularly updated, although non-representative, digital census that could fill data gaps. 

In addition to the limitations already discussed, current fertility data sources usually miss one side of the total picture: the male. Only a few developed countries produce statistics for male fertility through civil registration and vital statistics systems (CRVS) \cite{schoumaker_measuring_2017}. Male fertility has recently started to be analysed for Germany and Greece \cite{dudel_estimating_2016,tragaki_male_2014}, but also for developing countries with Demographic and Health Surveys \cite{schoumaker_measuring_2017}.  Facebook can provide information in regards to this yet under-analyzed aspect. 

%
%
\section{Related Work}

Demography, being a data driven discipline among the social sciences, is one of the fields of research which can benefit from the abundance of Internet data. The focus of demography is on fertility, mortality and migration. Migration studies, in comparison to studies of fertility and mortality, are benefiting the most from social media data. Indeed, migration has been studied through Yahoo, Twitter, LinkedIn and Facebook \cite{zagheni_you_2012,state_mesh_2013,state_migration_2014,zagheni_leveraging_2017}. The use of social media data in migration studies can shed new light on a branch of research in which the data available contains many deficiencies. 
Fertility research through Internet data has manifested itself into studies of fertility by age and location in the US \cite{ojala_fertility_2017}, in seasonality of mating-related Web searchers and consequential fertility \cite{markey_seasonal_2013}.  \cite{hitsch_matching_2010}, \cite{bellou_impact_2015}, and \cite{billari_internet_2016} focus on the impact of the diffusion of internet on the postponement in timing of marriages and births. The area of fertility desires and intentions have also been explored through Twitter \cite{adair_babyfever:_2014}. 
It is interesting that the area of mortality research is also employing social media data. In fact, the Web is rich with websites of family trees and these have been used to study life expectancy \cite{fire_data_2015}. The study of causes of death and the health of populations has also been affected by the Internet \cite{gittelman_new_2015}.

%
%
\section{Facebook Advertising Data}
Facebook's advertising platform  permits advertisers to selectively show their advertisement to Facebook users matching criteria specified by the advertiser. The criteria are divided into four sections of variables: (i) demographics, (ii) location, (iii) interests, and (iv) behaviour. The demographic section contains information based on traits such age, gender, relationship status, education, workplace, job titles and more. Under this section, we can find information concerning parenthood. Parents on Facebook are further sub-divided by the age of their children. Facebook allows advertisers to target individuals by their location from a country level, to a neighbourhood level. This information can be disaggregated further between individuals that are travelling and whom are living in the specific area. Through the section on interests, advertisers can target individuals based on their interests, which are inferred  based on pages liked and other signals. The behaviours section provides information concerning purchase behaviours, device usage and other activities.  It is worth noting that Facebook gathers these estimations based on information from sites other than just facebook.com, as long as those sites have a Facebook Like or share functionality, which means a connection to facebook.com. 

%
%
\section{Dataset}

Our research focuses on individuals with a Facebook profile, aged between 15 and 49 years old. The dataset was collected on January 2, 2018. Facebook's advertising platform does not provide data for Cuba, Iran, North Korea, Syria, and Sudan.
We are interested in women and men of a reproductive age, between 15-49 years old, who had a child in the last 12 months. Facebook already prepares an aggregated estimate for this category. The age variable is divided into 5 years gaps (15-19, 20-24,25-29, 30-34, 35-39, 40-44, 45-49).  We downloaded information about the total population of women and men in each age group as a proxy for exposure population. We used the Facebook Application Programming Interface (API) to download the data \cite{araujo_using_2017}. 

In addition to the above online data, we obtained United Nations estimates for fertility as ``ground truth'' data with which to compare the Facebook data.

%
%
\section{Measures}
Fertility analysis measures are generally computed considering one sex only, the female population. The reason for this standard is that only women in their reproductive ages can give birth. Total Fertility Rate (TFR) and Mean Age at Childbearing (MAC) are the typically computed measures. In this paper, we are not interested in creating a model for studying the distortion linked to the digital divide, Facebook penetration \cite{fatehakiaetal18wd} or language of the TFR. 
Our focus is on the MAC, which is computed separately for both  sexes. The Spearman correlation coefficient and the Mean Absolute Percentage Error are the measures used for comparing the Facebook estimates to the ground truth data. To guard against overfitting, i.e.\ overly optimistic estimates, we used a leave-one-out-cross-validation (LOOCV): one observation is removed from the sample, while the remaining observations are used for training a model and predict the value for the removed observation. 
This approach is repeated as many times as the number of observations for calculating the average correlation and MAPE by continents.

\subsection{Mean Age at Childbearing:} The Mean Age at Childbearing (MAC) is computed as the sum of the Age Specific Fertility Rate (ASFR) weighted by the mid-point of each age group, divided by the sum of the ASFR. MAC can be computed as follows: 

$$_{n}\textrm{F}_{x}[0,T]=\frac{\mbox{Births between 0 and T to women aged x to x+n}}{\mbox{Women risk having a child in the period 0 to T}}$$

$$MAC=\frac{\sum\limits_{x=15}^{49}(x+0.5) \times _{n}\textrm{F}_{x}}{\sum \limits_{x=15}^{49}( _{n}\textrm{F}_{x})}$$

Where $(x+0.5)$ is the mid-point for each age interval and $(x+0.5)\times _{n}\textrm{F}_{x}$ is the age-specific fertility rate for women or men whose age corresponds to the age group of which $(x+0.5)$ is the mid-point.

%
%
\section{Results}

%
%
\subsection{Mean Age at Childbearing}
The calculations of the Mean Age at Childbearing (MAC) for females and males are presented in the next two sections. We have not included in the analysis those countries in which Facebook Advertising Data reports user count estimates equal to 20, which is the default lower bound response, as 0 is never reported.

\subsubsection{Female}
The correlation of the Female MAC with the corresponding UN estimates is 0.47 ($p = 4.02e-08$). The calculation is made on 138 countries out of 194 available on the UN data \footnote{Countries for which Facebook reported the default result, indistinguishable from no users, were not included in the analysis.}. Dividing the result by continents, we can see that the correlation is negative for Africa and South America, while for the other continents it is positive. The highest correlation is obtained in Europe. 

\begin{table}[ht]
\centering
\caption{Spearman Correlation and MAPE for Female MAC by continent and by LOOCV.}
\label{tab:female_mac}\footnotesize
\begin{tabular}{l|lr|lr|r}
 &  \multicolumn{2}{|c}{Female MAC}
 & \multicolumn{2}{|c|}{LOOCV} \\
Continent & Correlation & MAPE & Correlation & MAPE & N\\ \hline \hline
Africa & -0.27* & 6.64\% & 0.11 & 3.06\% & 28 \\ \hline
Asia & 0.52*** & 6.80\% & 0.46*** & 3.92\% & 33\\ \hline
Europe & 0.69*** & 6.54\% & 0.66*** & 2.98\% & 41\\ \hline
North & 0.62*** & 6.52\% & 0.55*** & 3.15\% & 16\\
America & &  & & & \\ \hline
Oceania & 0.29 & 4.98\% & -0.47 & 3.80\% & 8 \\\hline
South & -0.26 & 6.88\% & -0.62** & 2.77\% & 12 \\
America & & & & &  \\\hline
\end{tabular}
\begin{flushleft} \tiny{*p<0.1; **p<0.05; ***p<0.01} \end{flushleft}
\end{table}

\subsubsection{Male}
The correlation for Male MAC is 0.79 ($p = 2.2e-15$). The calculation has been made on 82 countries out of 164 available countries in the UN data with the latest available estimates for the period 2006-2015. Only three African countries are included in this calculations. We calculated the correlation for Female MAC approximately for the same sample (71 countries) and the correlation is equal to 0.75 ($p = 6.21e-14$).

\begin{table}[ht]
\centering
\caption{Spearman Correlation and MAPE for Male MAC by continent and by LOOCV.}
\label{tab:male_mac}\footnotesize
\begin{tabular}{l|lr|lr|r}
 &  \multicolumn{2}{|c}{Male MAC}
 & \multicolumn{2}{|c|}{LOOCV} \\
Continent & Correlation & MAPE & Correlation & MAPE & N \\ \hline \hline
Africa & 1.00 & 6.44\% & 0.5 & 6.74\% & 3 \\ \hline
Asia & 0.75*** & 6.88\% & 0.46*** & 3.92\% & 17 \\ \hline
Europe & 0.71*** & 5.35\% & 0.70*** & 2.41\% & 40 \\ \hline
North  & 0.87*** & 2.01\% & 0.81***& 5.89\% & 13 \\
America & &  & & & \\ \hline
Oceania & 0.50 & 4.82\% & -0.5 & 3.25\% & 3 \\ \hline
South & 0.08 & 6.51\% & -0.48 & 3.63\% & 6 \\
America & & & & &  \\\hline
\end{tabular}
\begin{flushleft} \tiny{*p<0.1; **p<0.05; ***p<0.01} \end{flushleft}
\end{table}

\subsection{Modeling male fertility}

We fit a simple linear regression to model male fertility: 
$$MAC_{UN}= \beta_0 + \beta_1 MAC_{FB} + \epsilon$$
where $MAC_{UN}$ is the MAC calculated with the United Nation data, and $MAC_{FB}$
is the MAC estimated through Facebook's Advertising data. In Table \ref{tab:regression}, we report the result of the linear regression model. The R$^2$ (0.676) indicates a good fit of the model. On average, Facebook data is underestimating male MAC. 

\begin{table}[ht]
\centering
\caption{Linear Regression for Male MAC.}
\label{tab:regression}
\begin{tabular}{l|r}
                    & Estimate  \\ \hline \hline
Intercept           & 7.451***   (1.936)      \\
Facebook MAC        & 0.811***  (0.063)      \\
N                   & 81                 \\
R$^2$           & 0.676              \\
Adjusted R$^2$  & 0.671              \\
Residual Std. Error & 0.949    (df=79)    \\
F Static            & 164.4***  (df=1;79)\\
\hline
\end{tabular}
\begin{flushleft} \tiny{*p<0.1; **p<0.05; ***p<0.01} \end{flushleft}
\end{table}

To validate our results, we performed an out-of-sample exercise. The simple linear regression has been run ten times, each time including 72 randomly selected observations as training data and 10 as a test data set.  The average value of the MAPE for the predictions on the test set is equal to 2.3\%, indicating that the model has high predictive capacity. For reference, the (standard deviation)/(average) is 11.72\%.

\begin{figure}[ht]
\centering
  \includegraphics[height=6.5cm]{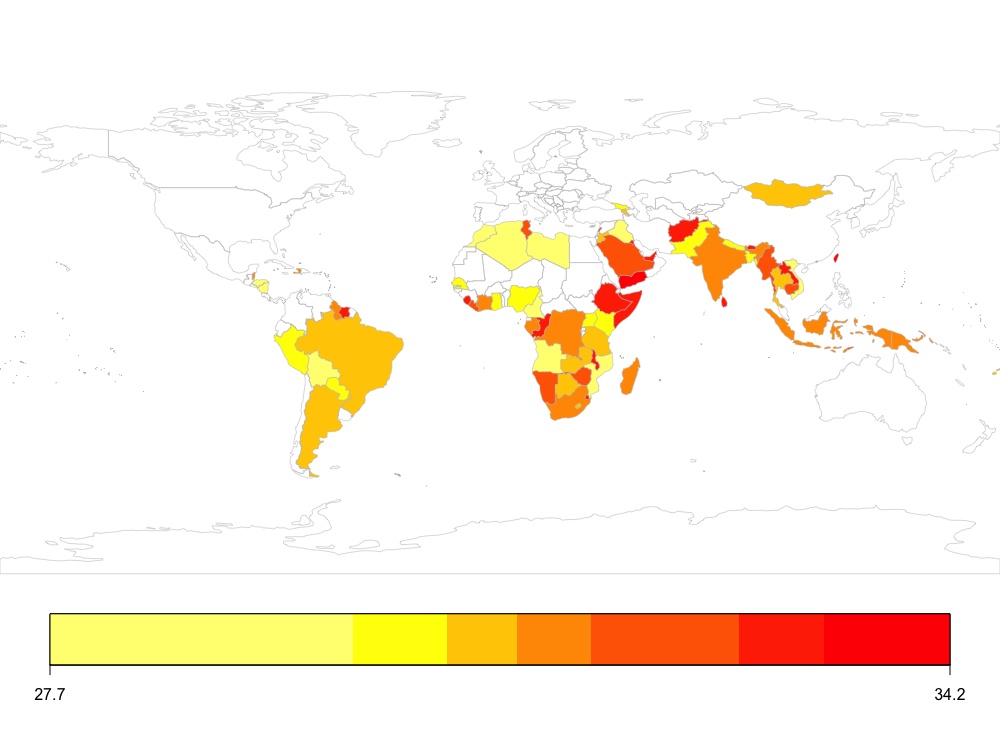}
 \caption{Prediction of Male MAC 2017 for countries without UN ground truth data.  }\label{men}
\end{figure}

Then, we predicted the Male MAC through a linear regression for those countries (79) without ground truth data. The prediction results are shown in Figure \ref{men}. As the map shows, our method helps to fill ``data gaps'' in many developing countries where these kinds of estimates are currently unavailable, potentially having big implications for policy making.

%
%
%

%
%
\section{Conclusion and Discussion}

This paper provides the basis for running more detailed  (male) fertility analysis through Facebook Advertising Data as it shows the feasibility to estimate Mean Age at Childbearing (MAC). Our work highlights the limitations as well as the advantages of this data source. 
One advantage of Facebook data is that MAC estimates can be produced instantaneously, in particular for under-studied dimensions such as male fertility. There are further possibilities with this data combining the analysis of fertility with other targeting variables provided by Facebook, such as education, relationship status or interests in certain topics, such as religious content. We believe that this is a promising and new direction for future work on more multi-faceted fertility research at a global scale. 
Another advantage is that, due to Facebook's global reach, we can study fertility in developing countries. This has shown promising results for certain countries and is a promising starting point to fill data gaps for countries where ground truth data are missing or not up-to-date. Moreover, the Internet penetration rate is increasing in these countries and globally, which will lead to more Internet users and therefore, more data obtainable through Facebook or other online advertising platforms. 

\section{Acknowledgement}
We thank the three anonymous reviewers for comments which improved the paper. We would also like to thank the European Doctoral School of Demography 2016-17 for the support and feedback on this research, especially Alyce Raybould for the help with English.

\bibliographystyle{aaai}
\bibliography{Fertility}{} 

\begin{thebibliography}{}

\bibitem[\protect\citeauthoryear{AbouZahr \bgroup et al\mbox.\egroup
  }{2015}]{abouzahr_civil_2015}
AbouZahr, C.; de~Savigny, D.; Mikkelsen, L.; Setel, P.~W.; Lozano, R.; Nichols,
  E.; Notzon, F.; and Lopez, A.~D.
\newblock 2015.
\newblock Civil registration and vital statistics: progress in the data
  revolution for counting and accountability.
\newblock {\em The Lancet} 386(10001):1373--1385.

\bibitem[\protect\citeauthoryear{Adair \bgroup et al\mbox.\egroup
  }{2014}]{adair_babyfever:_2014}
Adair, L.~E.; Brase, G.~L.; Akao, K.; and Jantsch, M.
\newblock 2014.
\newblock \#babyfever: {Social} and media influences on fertility desires.
\newblock {\em Personality and Individual Differences} 71:135--139.

\bibitem[\protect\citeauthoryear{Ara{\'{u}}jo \bgroup et al\mbox.\egroup
  }{2017}]{araujo_using_2017}
Ara{\'{u}}jo, M.; Mejova, Y.; Weber, I.; and Benevenuto, F.
\newblock 2017.
\newblock Using facebook ads audiences for global lifestyle disease
  surveillance: Promises and limitations.
\newblock In {\em WebSci},  253--257.

\bibitem[\protect\citeauthoryear{Bellou}{2015}]{bellou_impact_2015}
Bellou, A.
\newblock 2015.
\newblock The impact of {Internet} diffusion on marriage rates: evidence from
  the broadband market.
\newblock {\em Journal of Population Economics} 28(2):265--297.

\bibitem[\protect\citeauthoryear{Billari and Zagheni}{2017}]{billari_big_2017}
Billari, F.~C., and Zagheni, E.
\newblock 2017.
\newblock Big {Data} and {Population} {Processes}: {A} {Revolution}?
\newblock {\em SocArXiv}.

\bibitem[\protect\citeauthoryear{Billari}{2016}]{billari_internet_2016}
Billari, F.~C.
\newblock 2016.
\newblock Internet and the {Timing} of {Births}.
\newblock In {\em Giornate di {Studio} sulla {Popolazione} 2017}.

\bibitem[\protect\citeauthoryear{Bongaarts and
  Blanc}{2015}]{bongaarts_estimating_2015}
Bongaarts, J., and Blanc, A.~K.
\newblock 2015.
\newblock Estimating the current mean age of mothers at the birth of their
  first child from household surveys.
\newblock {\em Population Health Metrics} 13:25.

\bibitem[\protect\citeauthoryear{Dudel and
  Klüsener}{2016}]{dudel_estimating_2016}
Dudel, C., and Klüsener, S.
\newblock 2016.
\newblock Estimating male fertility in eastern and western {Germany} since
  1991: {A} new lowest low?
\newblock {\em Demographic Research} 35(53):1549--1560.

\bibitem[\protect\citeauthoryear{Fatehkia, Kashyap, and
  Weber}{}]{fatehakiaetal18wd}
Fatehkia, M.; Kashyap, R.; and Weber, I.
\newblock Using facebook ad data to track the global digital gender gap.
\newblock {\em World Development} 107:189--209.

\bibitem[\protect\citeauthoryear{Fire and Elovici}{2015}]{fire_data_2015}
Fire, M., and Elovici, Y.
\newblock 2015.
\newblock Data {Mining} of {Online} {Genealogy} {Datasets} for {Revealing}
  {Lifespan} {Patterns} in {Human} {Population}.
\newblock {\em ACM TIST} 6(2):28:1--28:22.

\bibitem[\protect\citeauthoryear{Gittelman \bgroup et al\mbox.\egroup
  }{2015}]{gittelman_new_2015}
Gittelman, S.; Lange, V.; Gotway~Crawford, C.~A.; Okoro, C.~A.; Lieb, E.;
  Dhingra, S.~S.; and Trimarchi, E.
\newblock 2015.
\newblock A {New} {Source} of {Data} for {Public} {Health} {Surveillance}:
  {Facebook} {Likes}.
\newblock {\em Journal of Medical Internet Research} 17(4).

\bibitem[\protect\citeauthoryear{Hitsch, Hortaçsu, and
  Ariely}{2010}]{hitsch_matching_2010}
Hitsch, G.~J.; Hortaçsu, A.; and Ariely, D.
\newblock 2010.
\newblock Matching and {Sorting} in {Online} {Dating}.
\newblock {\em The American Economic Review} 100(1):130--163.

\bibitem[\protect\citeauthoryear{Markey and
  Markey}{2013}]{markey_seasonal_2013}
Markey, P.~M., and Markey, C.~N.
\newblock 2013.
\newblock Seasonal variation in internet keyword searches: a proxy assessment
  of sex mating behaviors.
\newblock {\em Archives of Sexual Behavior} 42(4):515--521.

\bibitem[\protect\citeauthoryear{Mikkelsen \bgroup et al\mbox.\egroup
  }{2015}]{mikkelsen_global_2015}
Mikkelsen, L.; Phillips, D.~E.; AbouZahr, C.; Setel, P.~W.; de~Savigny, D.;
  Lozano, R.; and Lopez, A.~D.
\newblock 2015.
\newblock A global assessment of civil registration and vital statistics
  systems: monitoring data quality and progress.
\newblock {\em The Lancet} 386(10001):1395--1406.

\bibitem[\protect\citeauthoryear{Moultrie \bgroup et al\mbox.\egroup
  }{2013}]{moultrie_tools_2013}
Moultrie, T.~A.; Dorrington, R.~E.; Hill, A.~G.; Hill, K.; Timaeus, I.~M.; and
  Zaba, B.
\newblock 2013.
\newblock Tools for {Demographic} {Estimation}.

\bibitem[\protect\citeauthoryear{Ojala \bgroup et al\mbox.\egroup
  }{2017}]{ojala_fertility_2017}
Ojala, J.; Zagheni, E.; Billari, F.~C.; and Weber, I.
\newblock 2017.
\newblock Fertility and its meaning: Evidence from search behavior.
\newblock In {\em ICWSM},  640--643.

\bibitem[\protect\citeauthoryear{Schoumaker}{2017}]{schoumaker_measuring_2017}
Schoumaker, B.
\newblock 2017.
\newblock Measuring male fertility rates in developing countries with
  {Demographic} and {Health} {Surveys}: {An} assessment of three methods.
\newblock {\em Demographic Research} 36(28):803--850.

\bibitem[\protect\citeauthoryear{State \bgroup et al\mbox.\egroup
  }{2014}]{state_migration_2014}
State, B.; Rodriguez, M.; Helbing, D.; and Zagheni, E.
\newblock 2014.
\newblock Migration of {Professionals} to the {U}.{S}.
\newblock In {\em SocInfo},  531--543.

\bibitem[\protect\citeauthoryear{State \bgroup et al\mbox.\egroup
  }{2015}]{state_mesh_2013}
State, B.; Park, P.; Weber, I.; and Macy, M.
\newblock 2015.
\newblock The mesh of civilizations in the global network of digital
  communication.
\newblock {\em PLOS ONE} 10(5):1--9.

\bibitem[\protect\citeauthoryear{Tragaki and Bagavos}{2014}]{tragaki_male_2014}
Tragaki, A., and Bagavos, C.
\newblock 2014.
\newblock Male fertility in {Greece}: {Trends} and differentials by education
  level and employment status.
\newblock {\em Demographic Research; Rostock} 31:137--159.

\bibitem[\protect\citeauthoryear{UNICEF}{2015}]{unicef_state_2015}
UNICEF.
\newblock 2015.
\newblock The {State} of the {World}’s {Children} 2014 {In} {Numbers}:
  {Every} {Child} {Counts}.

\bibitem[\protect\citeauthoryear{United~Nations and
  Social~Affairs}{2015}]{un_2015}
United~Nations, D. o.~E., and Social~Affairs, P.~D.
\newblock 2015.
\newblock World population prospects: The 2015 revision, methodology of the
  united nations population estimates and projections.
\newblock Working Paper No. ESA/P/WP.242.

\bibitem[\protect\citeauthoryear{United~Nations and
  Social~Affairs}{2017}]{un_2017}
United~Nations, D. o.~E., and Social~Affairs, P.~D.
\newblock 2017.
\newblock World population prospects: The 2017 revision.
\newblock II:Demographic Profile (ST/ESA/SER.A/400)(2).

\bibitem[\protect\citeauthoryear{Zagheni and Weber}{2012}]{zagheni_you_2012}
Zagheni, E., and Weber, I.
\newblock 2012.
\newblock You {Are} {Where} {You} e-{Mail}: {Using} e-{Mail} {Data} to
  {Estimate} {International} {Migration} {Rates}.
\newblock In {\em WebSci},  348--351.

\bibitem[\protect\citeauthoryear{Zagheni and
  Weber}{2015}]{zagheni_demographic_2015}
Zagheni, E., and Weber, I.
\newblock 2015.
\newblock Demographic research with non-representative internet data.
\newblock {\em International Journal of Manpower} 36(1):13--25.

\bibitem[\protect\citeauthoryear{Zagheni, Weber, and
  Gummadi}{2017}]{zagheni_leveraging_2017}
Zagheni, E.; Weber, I.; and Gummadi, K.
\newblock 2017.
\newblock Leveraging {Facebook}'s {Advertising} {Platform} to {Monitor}
  {Stocks} of {Migrants}.
\newblock {\em Population and Development Review} 43(4):721--734.

\end{thebibliography}
\end{document}